\documentclass[authoryear]{elsarticle}

\usepackage{amssymb}
\usepackage{subcaption}
\usepackage{multirow}
\usepackage{hyperref}

\begin{document}

\begin{frontmatter}



\title {Learning solutions of parametric Navier-Stokes with physics-informed neural networks}  

\author[1]{Mahdi Naderibeni\corref{cor1}}
\ead{m.naderibeni@tudelft.nl}
\affiliation[1]{organization={Pattern Recognition and Bio-informatics Group, Delft University of Technology},
            addressline={Van Mourik Broekmanweg 6}, 
            city={Delft},
%          citysep={}, % Uncomment if no comma needed between city and postcode
            postcode={2628 XE}, 
            country={the Netherlands}}

\author[1]{Marcel J.T. Reinders}
\ead{m.j.t.reinders@tudelft.nl}

\author[2]{Liang Wu}
\ead{Liang.wu@dsm-firmenich.com}
\affiliation[2]{organization={Science, Research and Innovation, dsm-firmenich},
            city={Delft},
%          citysep={}, % Uncomment if no comma needed between city and postcode
            country={the Netherlands}}

\author[1]{David M.J. Tax}%[<options>]
\ead{D.M.J.Tax@tudelft.nl}

\cortext[cor1]{Corresponding author}

\begin{abstract}
    We leverage Physics-Informed Neural Networks (PINNs) to learn solution functions of parametric Navier-Stokes Equations (NSE). Our proposed approach results in a feasible optimization problem setup that bypasses PINNs' limitations in converging to solutions of highly nonlinear parametric-PDEs like NSE. We consider the parameter(s) of interest as inputs of PINNs along with spatio-temporal coordinates, and train PINNs on generated numerical solutions of parametric-PDES for instances of the parameters. We perform experiments on the classical 2D flow past cylinder problem aiming to learn velocities and pressure functions over a range of Reynolds numbers as parameter of interest. Provision of training data from generated numerical simulations allows for interpolation of the solution functions for a range of parameters. Therefore, we compare PINNs with unconstrained conventional Neural Networks (NN) on this problem setup to investigate the effectiveness of considering the PDEs regularization in the loss function. We show that our proposed approach results in optimizing PINN models that learn the solution functions while making sure that flow predictions are in line with conservational laws of mass and momentum.  Our results show that PINN results in accurate prediction of gradients compared to NN model, this is clearly visible in predicted vorticity fields given that none of these models were trained on vorticity labels.
\end{abstract}

\begin{highlights}
    \item Optimization of PINN for parametric Navier-Stokes equations can be facilitated by providing available solutions as training data.
    \item Considering parameter(s) of interest as direct inputs of PINNs allows for fast prediction of the solution for new instances of the parameters without re-optimization.
    \item PINN' predictions conserve physical laws on the task of interpolation between solutions of Navier-Stokes equations.
\end{highlights}

\begin{keyword}
Fluid dynamics\sep  Machine learning \sep Navier-Stokes equations \sep Physics-informed Neural Network
% %% MSC codes here, in the form: \MSC code \sep code
% %% or \MSC[2008] code \sep code (2000 is the default)
\end{keyword}

\end{frontmatter}


\section{Introduction}
Mathematical modeling of fluid motion is an important problem in science and engineering as fluids play an essential role in modeling many physical phenomena. The governing dynamics of motion in fluids can be derived from conservational laws of mass and momentum. The Navier-Stokes Equations are a set of non-linear Partial Differential Equations (PDEs) that represent the chaotic time-dependent behavior in fluids. These equations are parametric in nature and result in solutions that are different for different physical properties of fluids or geometries of the flow field. While they are nearly impossible to solve analytically, their numerical solution becomes computationally too expensive for high speed and complex flow fields. Classical numerical methods for solving PDEs require the solution domain to be discretized into many small sub-domains, assuming that solution functions remain constant in such sub-domains. However, given that smallest features in fluids are in the order of micro meters, a proper discretization of the solution domain requires dividing the solution domains into similar scales. Such high resolution discretization will not only suffer from the accumulation of round-off errors in numerical iterative methods, but also lead to prohibitively expensive computations.

Recent advancements in Machine Learning (ML) have motivated a wave of research to improve and accelerate fluid flow simulations. These development can be categorized based on the extent of contribution of data-driven techniques in the overall model. Hybrid approaches embed ML modeling into the Computational Fluid Dynamics'(CFD) iterative solver. In this setup the ML model becomes a part of the CFD solver either by providing approximations for turbulence closure, replacing the most computational expensive parts of the iterative CFD solver, or correcting between coarse and fine resolutions \citep{vinuesa2021potential, FONT2021110199}. Such approaches have resulted in faster and more accurate fluid simulations compared to classical CFD, however, they are associated with iterative numerical integration that makes them impracticable for real time predictions of unseen flow conditions.

% \textbf{operator learning}\\
Amongst data driven strategies in literature for solving parametric-PDEs \citep{takamoto_pdebench_2023}, operator learning approaches have gained a lot of attention \citep{kovachki_neural_2023, lu_deeponet_2021, li_fourier_2021, li_physics_informed_2023, wang_learning_2021}. In these approaches the task is to learn PDEs as operators that perform a mapping between function spaces. Both parameterized boundary conditions and parameterized PDEs can be considered in this problem setup. However one of their major limitations is the need for large datasets of input-output functions, a constraint that makes them infeasible for the problem setups with limited data. While the task of operator learning approaches is to learn the mapping between input-output functions, an alternative approach for dealing with PDEs is to directly learn the mapping from spatio-temporal coordinates to the solution functions. 

% \textbf{Physics-Informed Neural Networks}\\
\citep{raissi2019physics} introduced Physics-Informed Neural Networks (PINNs), in which by leveraging automatic differentiation, PDEs or governing constraints are embedded into the loss function of a deep artificial neural network. PINNs have shown to be suitable for both forward and inverse problems and have the potential to tackle ill-posed problems \cite{cai2021physics}. Due to the flexibility of PINNs, the structure of the model or the optimization problem remains the same for all the cases. In forward problems, where the aim is to solve the intended PDEs given their boundary and initial conditions, researchers have used PINNs as a solver for various types of PDEs including Navier-Stokes Equations \citep{jin_nsfnets_2021, hennigh2021nvidia}. Despite the flexibility and capabilities of PINNs, their application becomes more challenging as the complexity of the PDEs of interest increases. Since their introduction by \citep{raissi2019physics}, many PINN variants have tried to improve its performance. Two main groups of challenges that highly affect the learning dynamics of PINNs are: spectral bias of neural networks, and training point sampling strategies.

\textbf{PINN architecture and spectral bias of neural networks}. ~\citep{rahaman2019spectral} showed that deep neural networks are biased towards learning low frequency functions (less complex functions that vary globally without local fluctuations), to articulate it differently, over-parameterized networks prioritize learning simple patterns that generalize across data samples. This phenomenon adversely affects the training of PINNs and asks for small learning rates and long training procedures ~\citep{wang2022and}.  Fourier feature maps \citep{tancik2020fourier} and Sinusoidal Representation Networks (SIRENs) \cite{sitzmann2020implicit} are steps taken toward alleviating this problem \citep{wang2021eigenvector}. Fourier feature maps transform the input to a high dimensional feature space of high frequency functions and SIRENs use periodic activation functions. 

\textbf{Multi-objective loss function and Sampling strategies}. Given that the loss function in a PINN model is composed of several terms (at least two: one for the error between exact and predicted function values and one for PDEs residuals), each of these terms might act differently during training and the priority of the optimizer to minimize each term might change because of the different nature of the loss terms. \citep{wang2022and} reported a discrepancy in convergence rate of different loss components and they performed a Neural Tangent Kernel (NTK) study of PINNs and suggested a method to assign adaptive weights to the loss components based on eigenvalues of the NTK. The discrepancy in convergence rate of terms in the loss function is a direct result of 
the selected strategy in sampling training points for PINNs. While using PINNs, there is a freedom to sample points arbitrarily from the specified boundary and domain. \citep{raissi2019physics} sampled boundary and domain's internal points from uniform distributions. The freedom in sampling training points have resulted in a lack of a universal sampling strategy that is optimal for all problems. This has lead some researches to investigate automatic approaches for sampling training points. For example, \citep{lu_deepxde_2021} proposed a Residual-based Adaptive Refinement (RAR) method, where more residual points are added to the regions that PDEs are not satisfied, these additional points are selected from a dense pool of points in the domain. Furthermore, \cite{nabian_efficient_2021} proposed an adaptive method to re-sample all the residual points from a probability density function proportional to PDEs residual.

Literature on the application PINNs for the problem of parametric-PDEs can be summarized into two main categories. The first category leverages meta-learning and transfer-learning concepts to accelerate PINNs optimization process by learning a mapping between the parameter(s) of interest and the trained weights of PINN ~\citep{liu_novel_2022, penwarden_metalearning_2023}. While such approaches are useful in speeding up the training of PINNs, they require an optimization for each new instance of parameter and become infeasible for cases for which a real time solution of PDEs is expected. The second category considers the parameter(s) of interest as inputs of PINN along with variables of the PDEs. ~\citep{hennigh2021nvidia} used a PINN model to find the optimal geometry configuration for a heat sink. They trained a conjugate heat transfer problem of turbulent flow over a 3D complex geometry and used the trained model to optimize the heat sink geometry. They compared the compute time of their approach with numerical solvers, concluding that PINNs significantly accelerates design optimization. In a similar approach, ~\citep{demo_extended_2021} used parameters as inputs to PINN models to prepare a setup for parametric optimal control of PDEs. The approach of considering parameters as inputs to the PINN model, inherits the challenges related to PINNs optimization and becomes a much more difficult task for highly nonlinear PDEs like Navier-Stokes equations. 

In this paper we focus on the task of learning the solution of parametric Navier-Stokes Equations, considering the Reynolds number as the parameter of interest. We assume that limited data is available from numerical solutions of these PDEs for a few parameter values. We aim to learn the general solution of parametric Navier-Stokes for a range of Reynolds numbers in order to perform fast predictions of flow for new parameter instances. Considering the 2D classical problem of a flow passing a cylinder, we learn the mapping between Re number along with three spatio-temporal coordinates and the solution functions (velocities and pressure). While optimizing a PINN model for learning a complex fluid flow system over a range of PDE parameters is challenging due to previously mentioned limitations, we try to bypass these limitations by providing data from numerical solutions of the PDE for some instances of the parameter setting. For this purpose, we simulate the flow with existing CFD solvers to generate training data for specific flow conditions and we provide these data in training of the PINN's deep Neural Network. Given that we are providing such data during training, we compare the performance of the PINN model with a conventional unconstrained Neural Network model that is trained to interpolate the solution functions between the provided parameters.\\

% \textbf{Contributions}. 
Summarizing, our primary contributions are:
\begin{itemize}
    \item [-] We propose a strategy for applying PINNs on problems of parametric PDEs, enabling a feasible optimization process and fast predictions of solution functions for new instances of the parameters.
    \item [-] We propose to bypass the limitations associated with the optimization of PINNs for parametric-PDEs by training on data from generated numerical solutions for a limited set of instances of the parameter.
    \item [-] We show that our proposed approach results in optimizing PINN models that learn the parametric solution functions of NSE while making sure that the fluid flow predictions are in line with conservational laws of mass and momentum.
    % \item [-] Our results show that PINN results in accurate prediction of gradients compared to NN model, this is clearly visible in predicted vorticity fields given that none of these models were trained on vorticity labels.
\end{itemize}

% \textbf{Organization of the paper}\\
% Section \ref{method} outlines a review of PINNs approach, there we explore its potentials and limitations. In section \ref{experiment_setup} we explain our experimental setup for investigating the proposed approach and section \ref{Results} we show our results on performance of the approach. Section \ref{Conclusions} contains our concluding remarks, including directions for further investigations.\\

% we investigate prediction of vorticity field leveraging automatic differentiation. 

\section{Physics-Informed Neural Networks}\label{method}
The task of a PINN is to approximate the solution function of PDEs with a deep neural network. This network performs the mapping between the domain coordinates and the solution function(s). Following the basic formulation of a PINN \cite{raissi2019physics}, we consider a nonlinear parameterized partial differential equation of a general form:
\begin{equation}
    \frac{\partial u}{\partial t} + \mathcal{N}\left [ u(x, t); \lambda \right ]=0, x\in \Omega, t\in \left [ 0, T \right ]
\end{equation}
In this equation $u(x, t)$ is the solution function, $x$ and $t$ are spatial and temporal coordinates, $\mathcal{N[.]}$ is a parameterized nonlinear differential operator, and $\Omega$ is the spatial domain ($\Omega \subset \mathrm{R}^{D}$). A deep neural network approximates $u(x, t)$ by training on two types of data points from the spatio-temporal domain. Labeled points from initial and boundary conditions $\left \{ x_{u}^{i}, t_{u}^{i}, u^{i} \right \}_{i=1}^{N_{u}}$, where the exact values of $u(x, t)$ is known, and domain internal points with no labels $\left \{ x_f^{i}, t_f^{i} \right \}_{i=1}^{N_{f}}$, initially referred to as collocation points, but here we refer to them as residual points. For residual points, although the label is not available, the PINNs approach allows for learning to satisfy the governing PDE. An essential part of the PINN algorithm is to compute the derivative terms present in the PDEs with automatic differentiation, later on these terms are rearranged to recreate the residual form of the PDE:
\begin{equation}
    f :=\frac{\partial u}{\partial t} + \mathcal{N}\left [ u(x, t); \lambda \right ]
\end{equation}
To embed the governing physics into the neural network, the $f(t, x)$ function is included in the networks loss function (Eq.\ref{eq:PINNs_mse}). The parameters of this network can be learned by minimizing the mean square error loss:
\begin{equation}\label{eq:PINNs_mse}
    Loss:= \frac{1}{N_u}\sum_{i=1}^{N_u}(\left | u(x_{u}^{i}, t_{u}^{i})-u^{i}) \right |^{2}           +\frac{1}{N_f}\sum_{i=1}^{N_f}(\left | f(x_{f}^{i}, t_{f}^{i}) \right |^{2}
\end{equation}

\section{Experiment setup; 2D Navier-Stokes Equations}\label{experiment_setup}
\begin{figure}
\centering
  \begin{minipage}{0.54\linewidth}
    \begin{subfigure}{0.475\linewidth}
      \includegraphics[width=\linewidth]{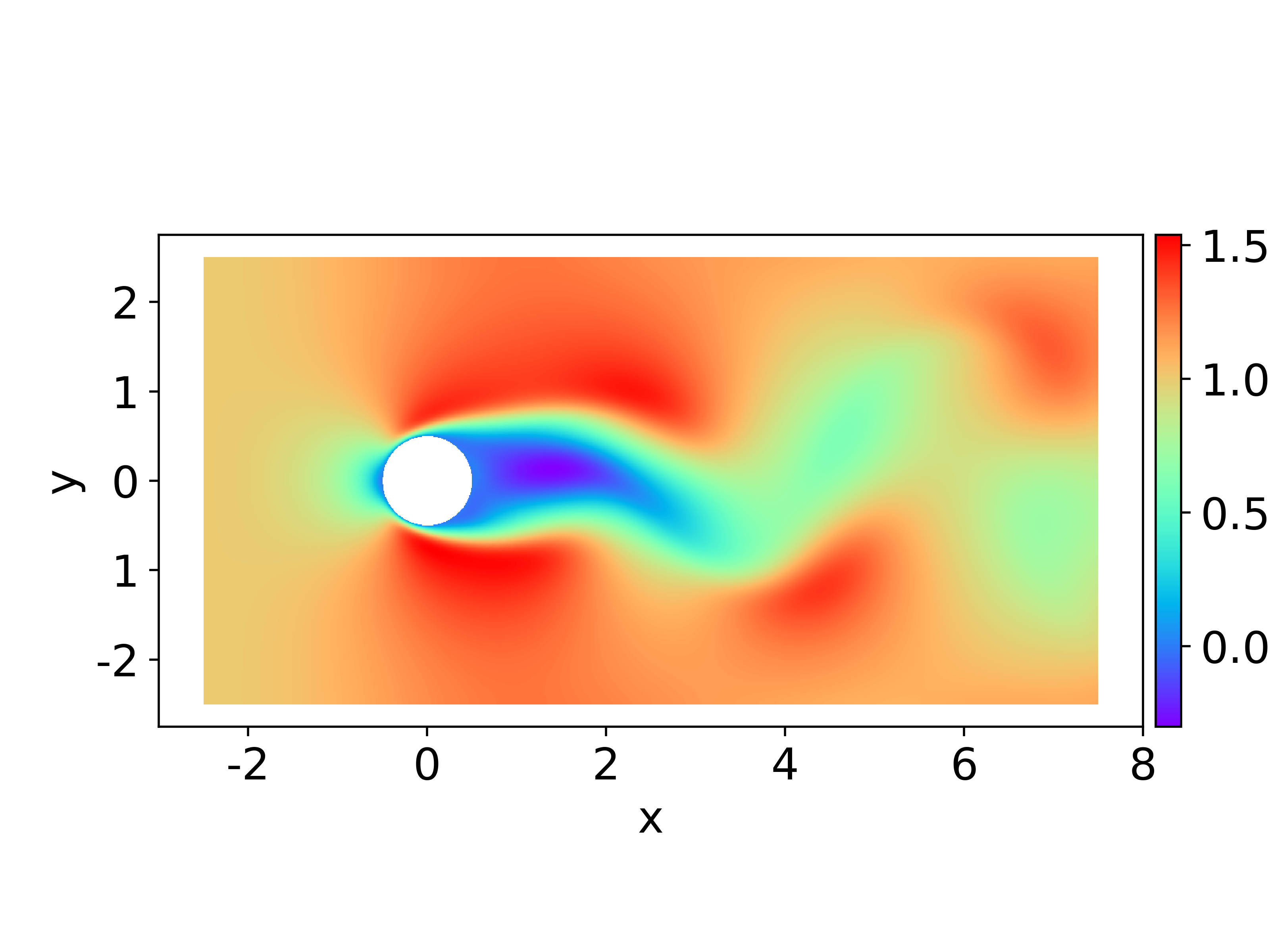}
      \caption{$\mathrm{Velocity_x}$; $Re=100$.}\label{subfig:x_velocity_Re100}
    \end{subfigure}
    \begin{subfigure}{0.475\linewidth}
      \includegraphics[width=\linewidth]{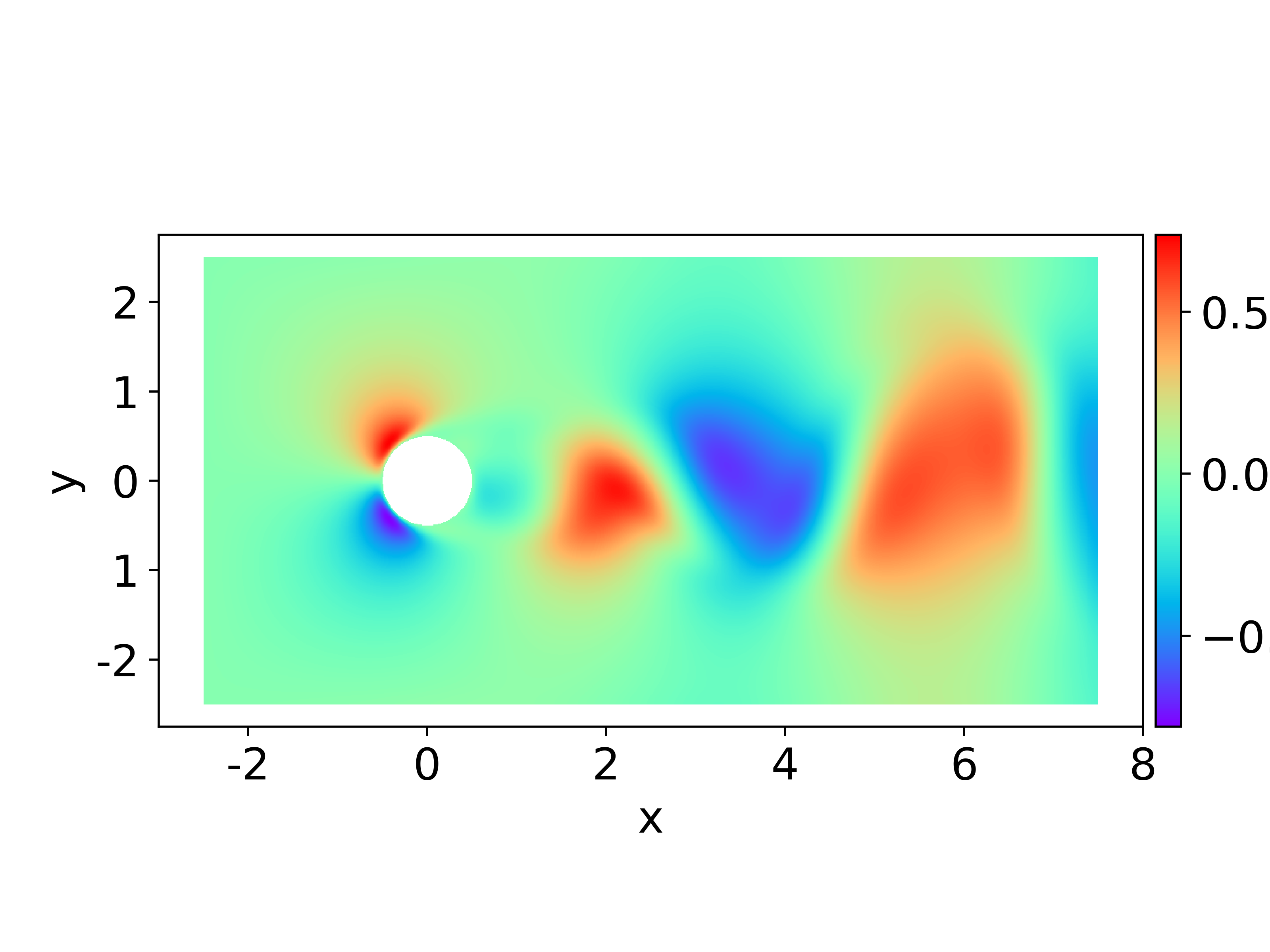}
      \caption{$\mathrm{Velocity_y}$; $Re=100$.}\label{subfig:y_velocity_Re100}
    \end{subfigure}

    \begin{subfigure}{0.475\linewidth}
      \includegraphics[width=\linewidth]{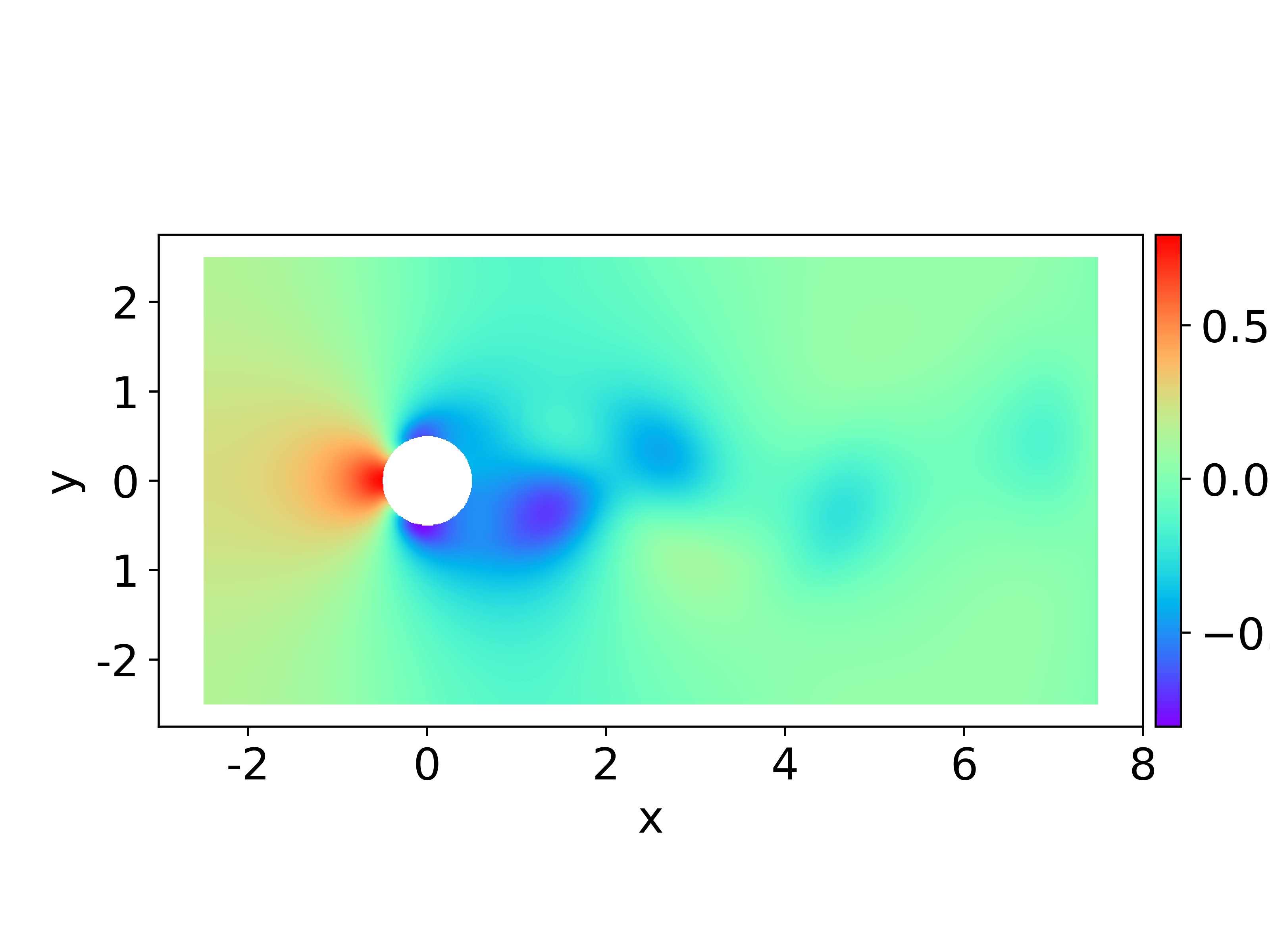}
      \caption{$\mathrm{Pressure}$; $Re=100$.}\label{subfig:pressure_Re100}
    \end{subfigure}
    \begin{subfigure}{0.475\linewidth}
      \includegraphics[width=\linewidth]{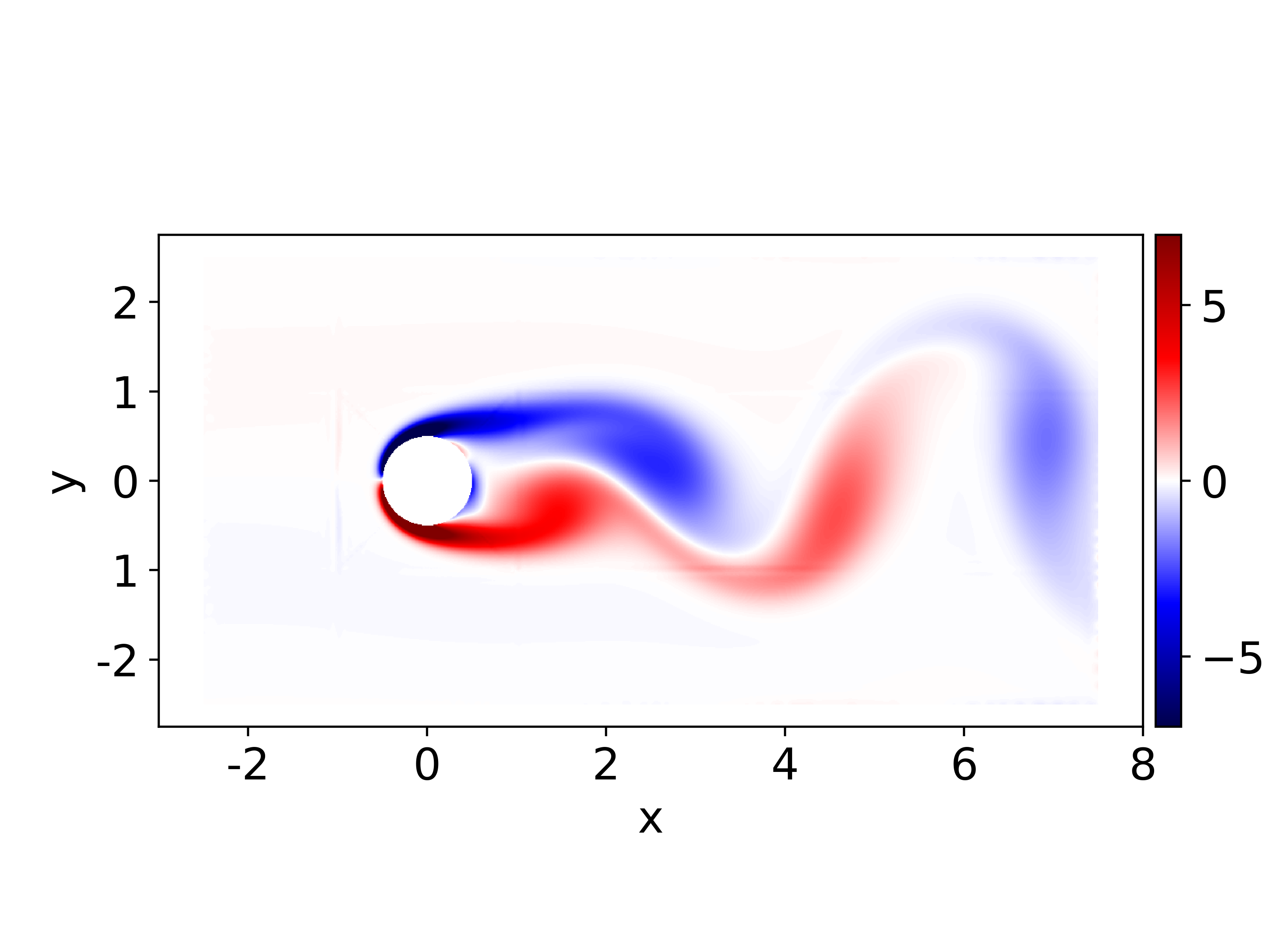}
      \caption{$\mathrm{Vorticity}$; $Re=100$.}\label{subfig:vorticity_Re100}
    \end{subfigure}
  \end{minipage}  
  \begin{minipage}{0.44\linewidth}
  \begin{subfigure}{\linewidth}
    \includegraphics[width=\linewidth]{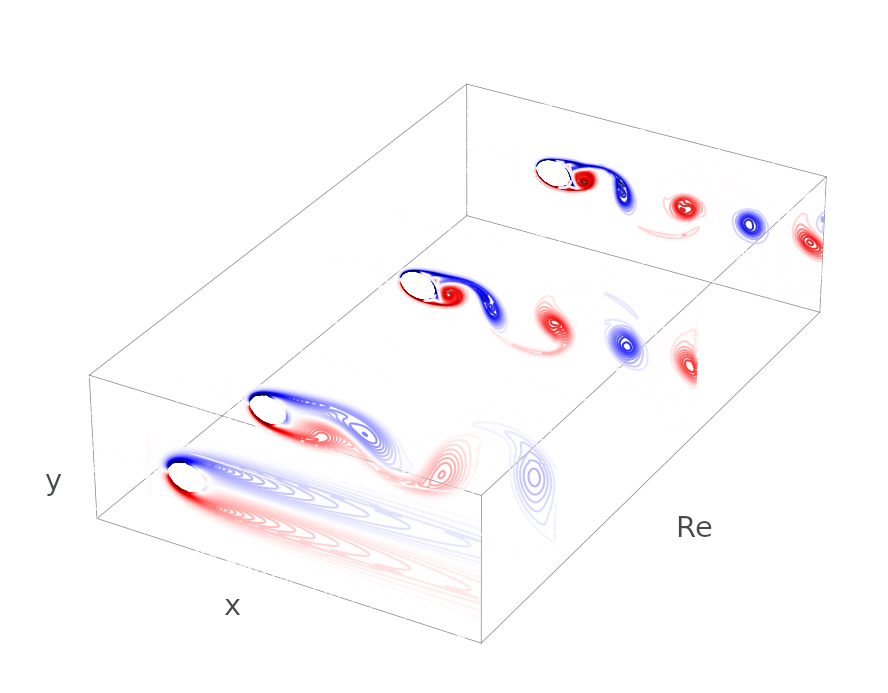}
    \caption{Vorticity versus Re.}\label{subfig:Re_vorticities}
  \end{subfigure}
  \end{minipage}
  \caption{ A snapshot of generated data for flow past cylinder problem for (a) the x component of velocity, (b) the y component of velocity, (c) pressure and (d) the vorticity (the curl of velocity field), generated for a flow with $Re=100$. Figure (e) demonstrates the evolution of a snapshot of vorticity with changing Re number.}
  \label{fig:uvpv_Res}
\end{figure}

To generate CFD data for the fluid flow past cylinder problem, we employed the OpenFoam CFD solver to simulate flow fields for several Re numbers (refer to appendix \ref{appendix:a}). The Re number represents the ratio between inertial forces to viscous forces within a fluid, it embeds information about the geometry, velocity and properties of the fluid into a dimensionless number. Fig. \ref{subfig:x_velocity_Re100}, \ref{subfig:y_velocity_Re100}, and \ref{subfig:pressure_Re100} show a snapshot of the developed velocity and pressure for a flow with $Re=100$. In this problem setup, while the flow remains unseparated at small Reynolds numbers ($Re<5$), by increasing the Re number, the flow develops an irrational periodicity in ($40<Re<150$). At higher Re Numbers, a transition to turbulent regime occurs. Fig. \ref{subfig:Re_vorticities} shows the evolution of vorticity (curl of velocity field) by changing the Re number. 

\textbf{Model setup and training}. To learn the general solution functions for a range of Re numbers, the Re number is considered as a direct input to the neural network along with the spatio-temporal coordinates. Therefore, the PINN model performs a mapping between $(x, y, t , Re)$ and the solution functions $(u, v, p)$. To construct the multi-task loss function of the PINN model, the residual forms of Navier-Stokes Equations are:
\begin{equation}
    f:= \frac{\partial u}{\partial x}+\frac{\partial v}{\partial y}
\end{equation}
\begin{equation}
    g:= \frac{\partial u}{\partial t} + (u\frac{\partial u}{\partial x}+v\frac{\partial u}{\partial y})+\frac{\partial p}{\partial x}    - \frac{1}{Re} (\frac{\partial^2 u}{\partial x^2}+\frac{\partial^2 u}{\partial y^2})
\end{equation}
\begin{equation}
    h:=\frac{\partial v}{\partial t} +  (u\frac{\partial v}{\partial x}+v\frac{\partial v}{\partial y})
        +\frac{\partial p}{\partial y} -\frac{1}{Re} (\frac{\partial^2 v}{\partial x^2}+\frac{\partial^2 v}{\partial y^2})
\end{equation}
In these equations $u$ and $v$ are components of 2D velocity vector field ($V=u\vec{i}+v\vec{j}$) and $p$ is the pressure (a scalar). 

\begin{figure}[h]
    \centering
    \includegraphics[width=0.8\textwidth]{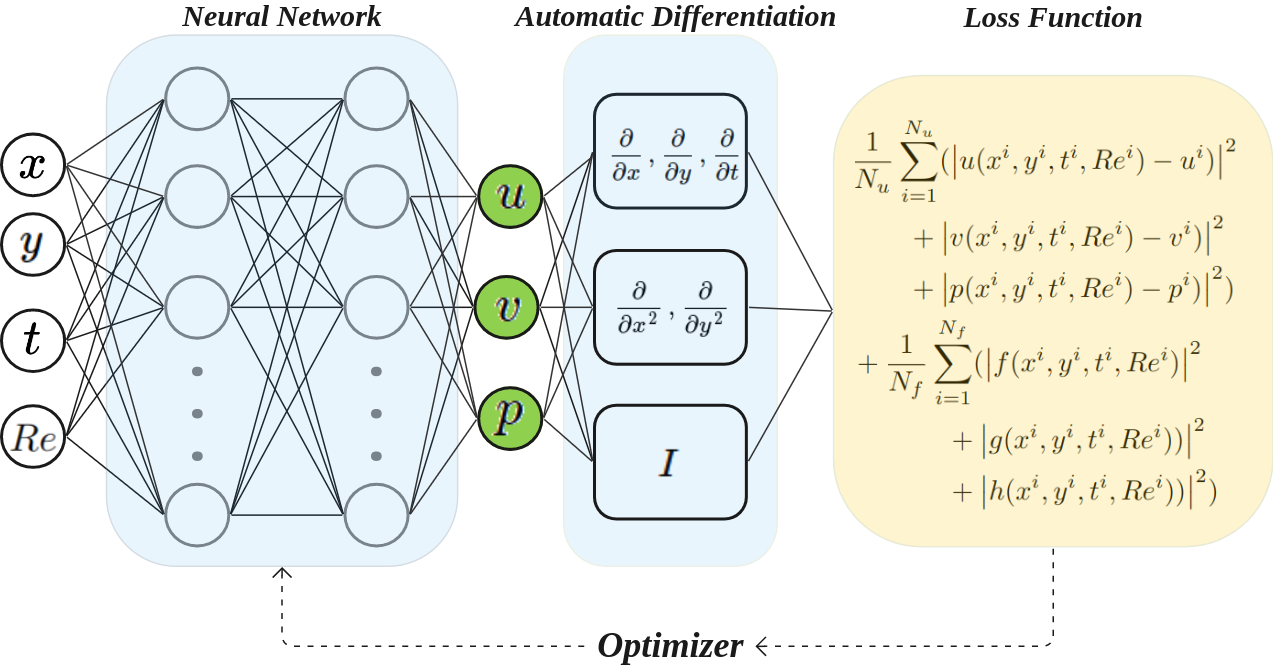}
    \caption{PINN architecture for solving Navier-Stokes Equation.}
    \label{fig:NS_NN}
\end{figure}

\begin{figure}[ht]
  \centering
    \begin{minipage}{0.49\linewidth}
    \centering
          \begin{subfigure}{\textwidth}
                \centering
                \includegraphics[clip, trim=0.01cm 3cm 0.01cm 3cm, width=0.7\textwidth]{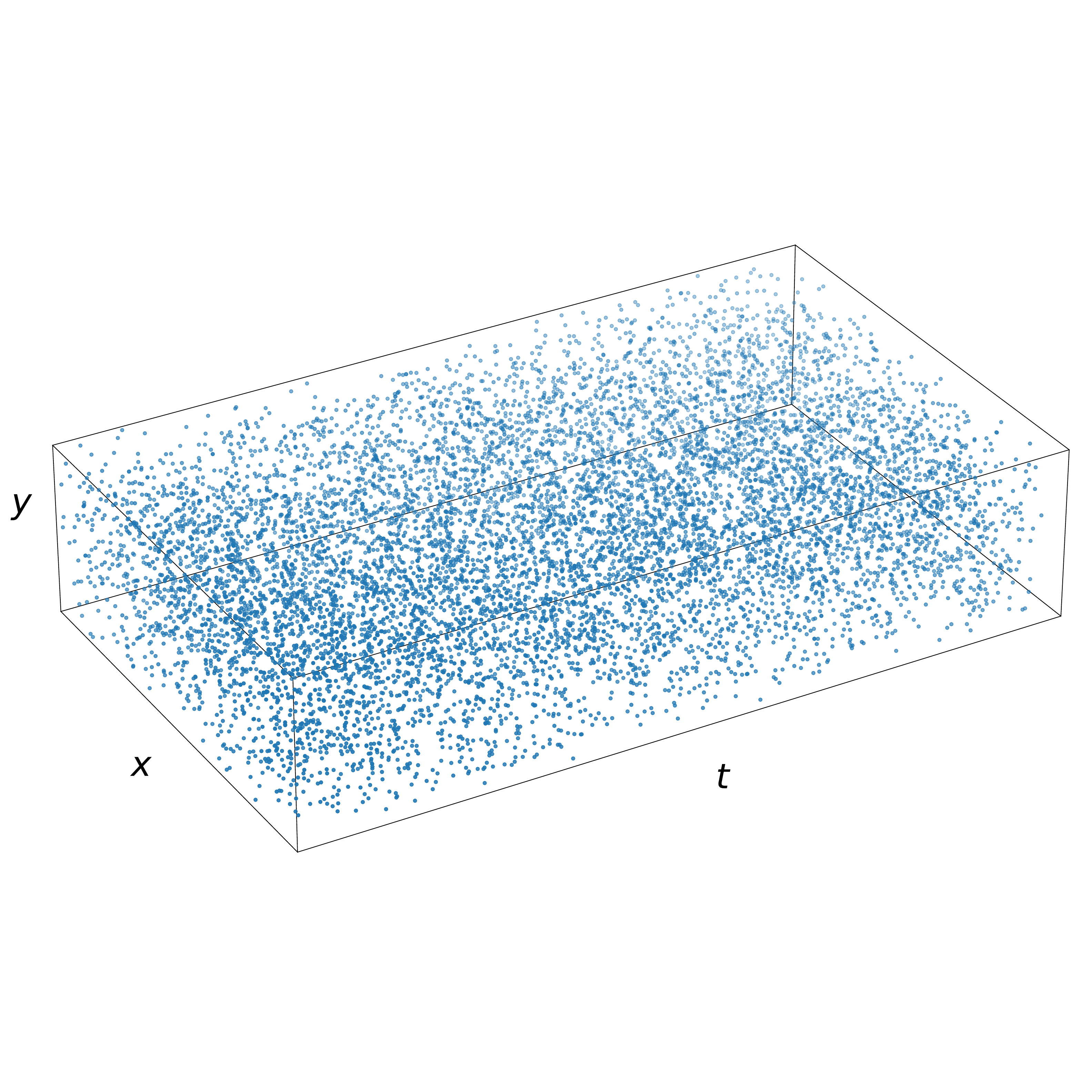}
                \caption{}%Uniform distribution of residual points}
                \label{subfig:uniform_lhs_sampling}
          \end{subfigure}
          \begin{subfigure}{\textwidth}
            \centering
               \includegraphics[width=0.7\textwidth]{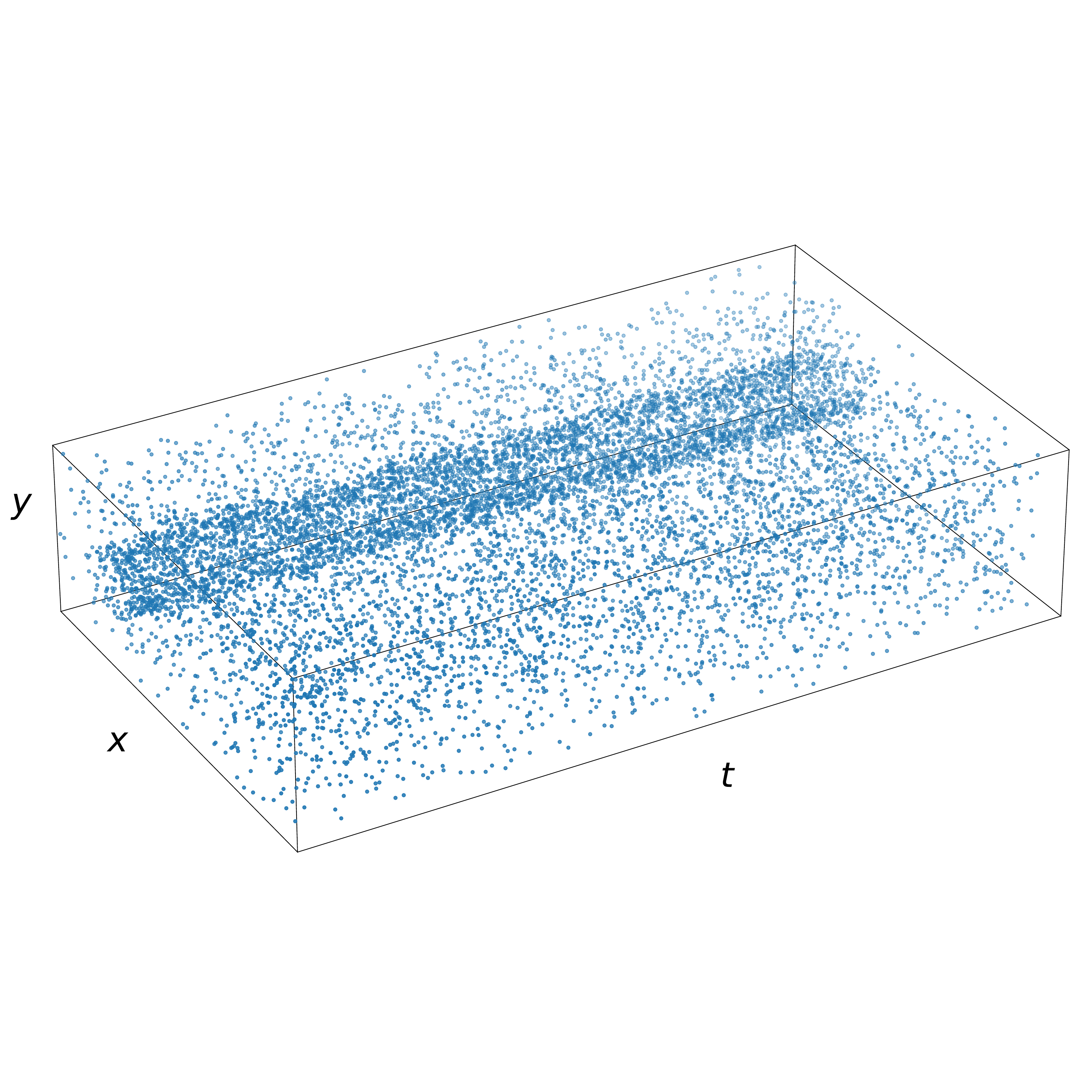}
                \caption{}%Uniform distribution of residual points that are refined around the cylinder.}
                \label{subfig:nonuniform_lhs_sampling}
          \end{subfigure}
  \end{minipage}
  % \hfill
  \begin{minipage}{0.49\linewidth}
  \centering
        \begin{subfigure}{\linewidth}
        \centering
            \includegraphics[width=0.7\linewidth]{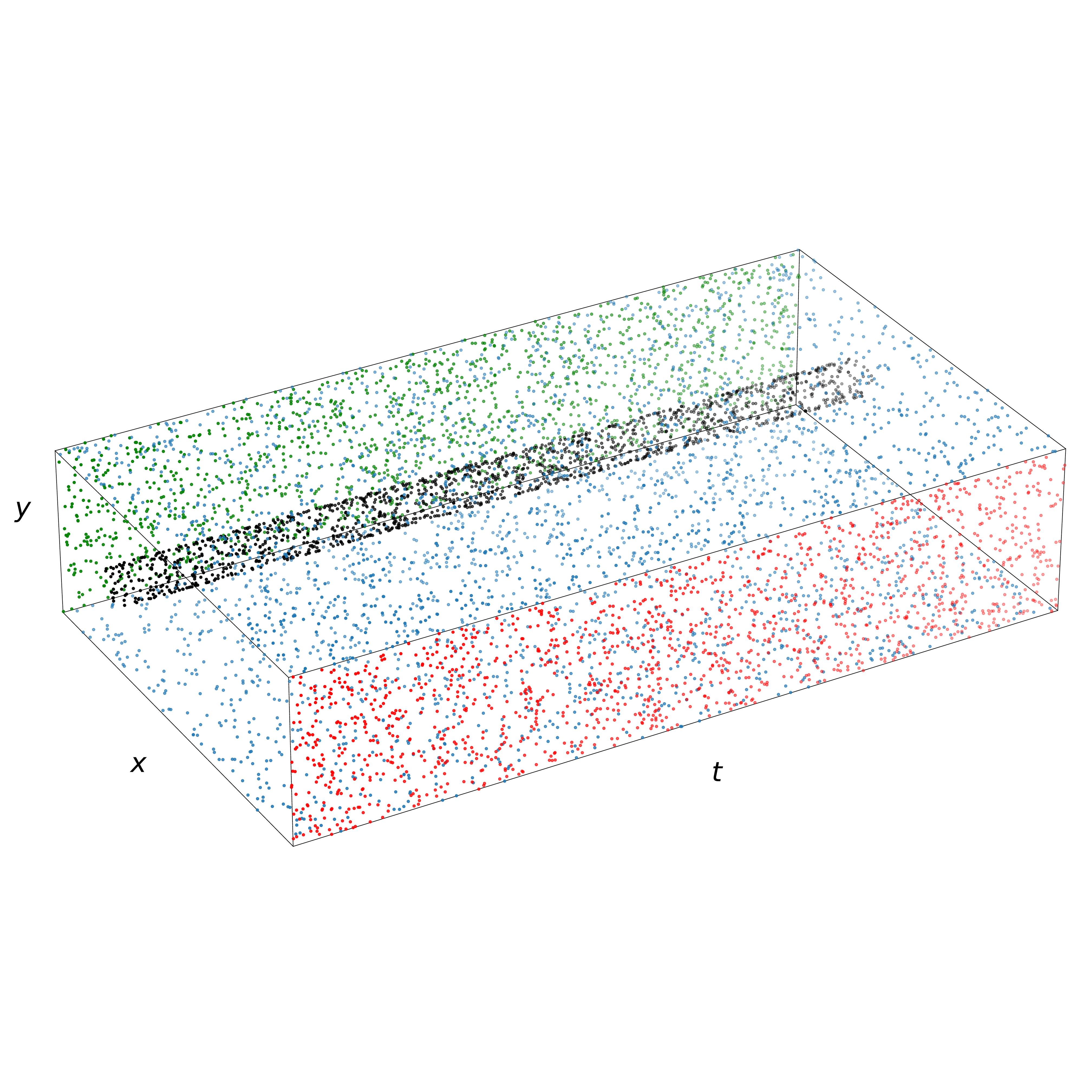}
            \caption{}%Initial and Boundary conditions points.}\label{subfig:Re_vorticities}
            \label{subfig:BC_points}
        \end{subfigure}
        \begin{subfigure}{\linewidth}
        \centering
            \includegraphics[width=0.7\linewidth]{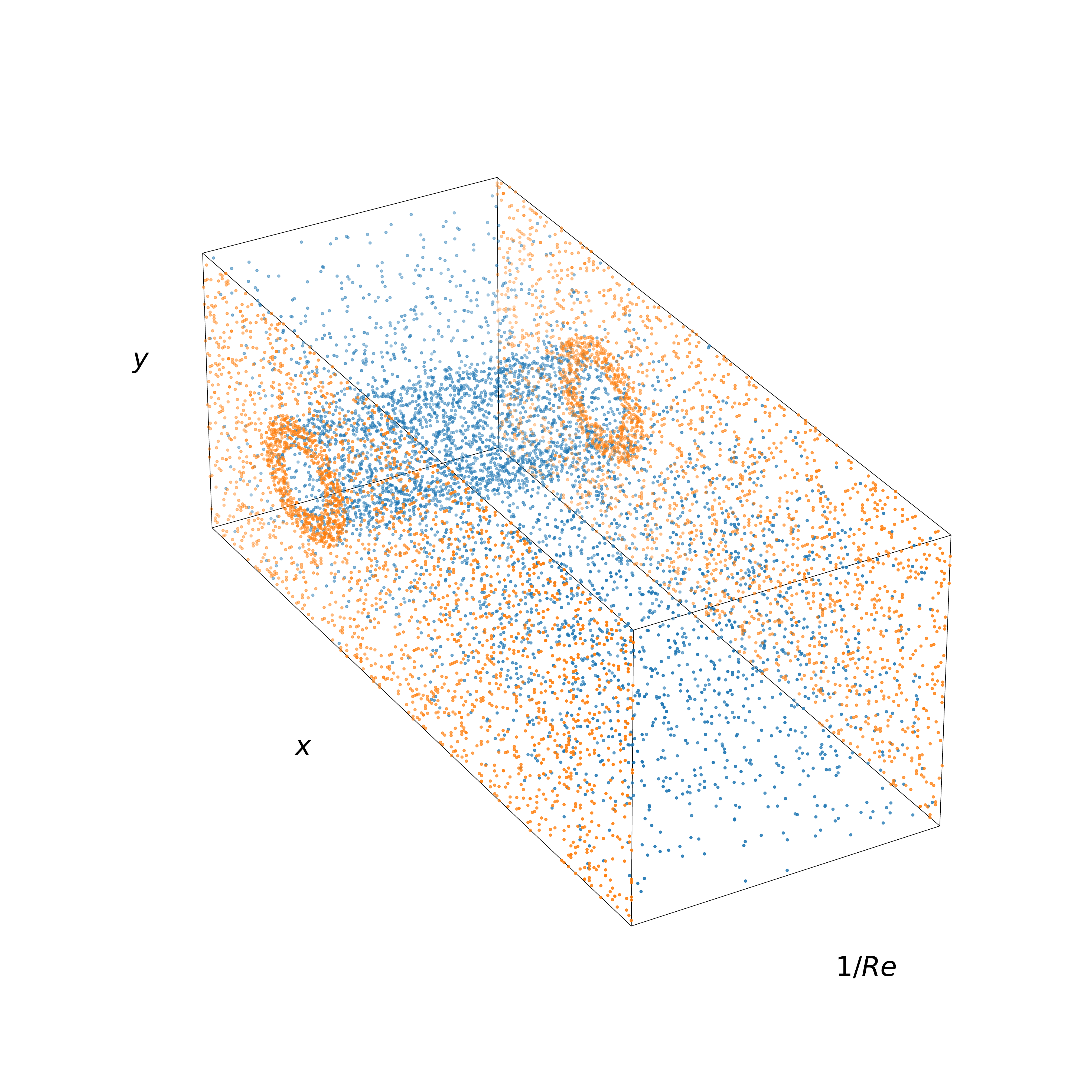}
            \caption{}%Distribution of points along the Re parameter.}\label{subfig:Re_vorticities}
        \end{subfigure}
  \end{minipage}  
  
  \caption{Strategies for sampling data points required in training PINN for the flow past cylinder problem. Residual points with (a) a uniform distribution, (b) a uniform distribution that is refined around the cylinder. (c) Training points sampled from initial and boundary conditions. (green : inlet, red: outlet, black: cylinder wall, blue: points on top and bottom). (d) Distribution of points along the Re parameter. (blue: residual points, orange: points from two available solutions)}
\end{figure}

The corresponding architecture for our PINN model is depicted in Fig. \ref{fig:NS_NN}. This model has a Fourier feature map with 50 bins, followed by a fully connected Neural network consisting of 7 hidden layers of 100 neurons each. The training dataset consists of $N_{u}=\mathrm{500k}$, labeled and $N_{f}=\mathrm{800k}$ residual data points that are sampled from uniform distributions along $(x, y, t, Re)$ coordinates. The labeled dataset includes $\mathrm{400k}$ points sampled from uniform distributions of CFD data, generated for $1/Re = [2, 2.5, 3, 5, 10] \times 10^{-3}$. The remaining $\mathrm{100k}$ labeled points are sampled from initial and boundary conditions (Fig. \ref{subfig:BC_points}). Based on this setup, two type of models are trained with and without the PDEs residuals term, to see the effectiveness of PINNs compared to Neural Networks. The Adam optimizer with a learning rate of $10^{-3}$ is used to adjust model parameters in $\mathrm{30k}$ epochs.

In the flow past cylinder problem, regions around the cylinder and the wake after it are under considerable variations of the flow field while other regions comparably remain stable (Fig. \ref{subfig:vorticity_Re100}). Sampling strategies that pay more attention to the difference between those regions are expected to provide a richer dataset for training PINNs. Our initial experiments showed that sampling points from a uniform distribution (Fig. \ref{subfig:uniform_lhs_sampling}) is not an efficient strategy to obtain acceptable solutions; i.e. a comparably large number of training points need to be sampled. Conversely, by putting a focus on the region around the cylinder, less data points are required (Fig. \ref{subfig:nonuniform_lhs_sampling}). In this problem setup, we assumed the flow domain size to be $[-2.5D, 7.5D]\times[-2.5D, 2.5D]$ where $D$ is the diameter of the cylinder located at $(0, 0)$. A uniform velocity flow ($u=1, v=0$) enters the domain (colored green in Fig. \ref{subfig:BC_points}) and passes over the cylinder with no-slip ($u=0, v=0$) condition on the wall (colored black in Fig. \ref{subfig:BC_points}), the outlet pressure (colored red in Fig. \ref{subfig:BC_points}) is zero ($p=0$) and periodic boundary condition is applied on top and bottom walls (colored blue in Fig. \ref{subfig:BC_points}).\\

\section{Results and Discussions}\label{Results}
In this section, we evaluate the flow field predictions of our PINN model compared to a conventional Neural Network (NN) model. These models were trained on data provided from fluid simulations for several Reynolds numbers in the range of $\mathrm{100}$ and $\mathrm{500}$.\\

\begin{figure}[h]
    \centering
    \includegraphics[width=0.7\textwidth]{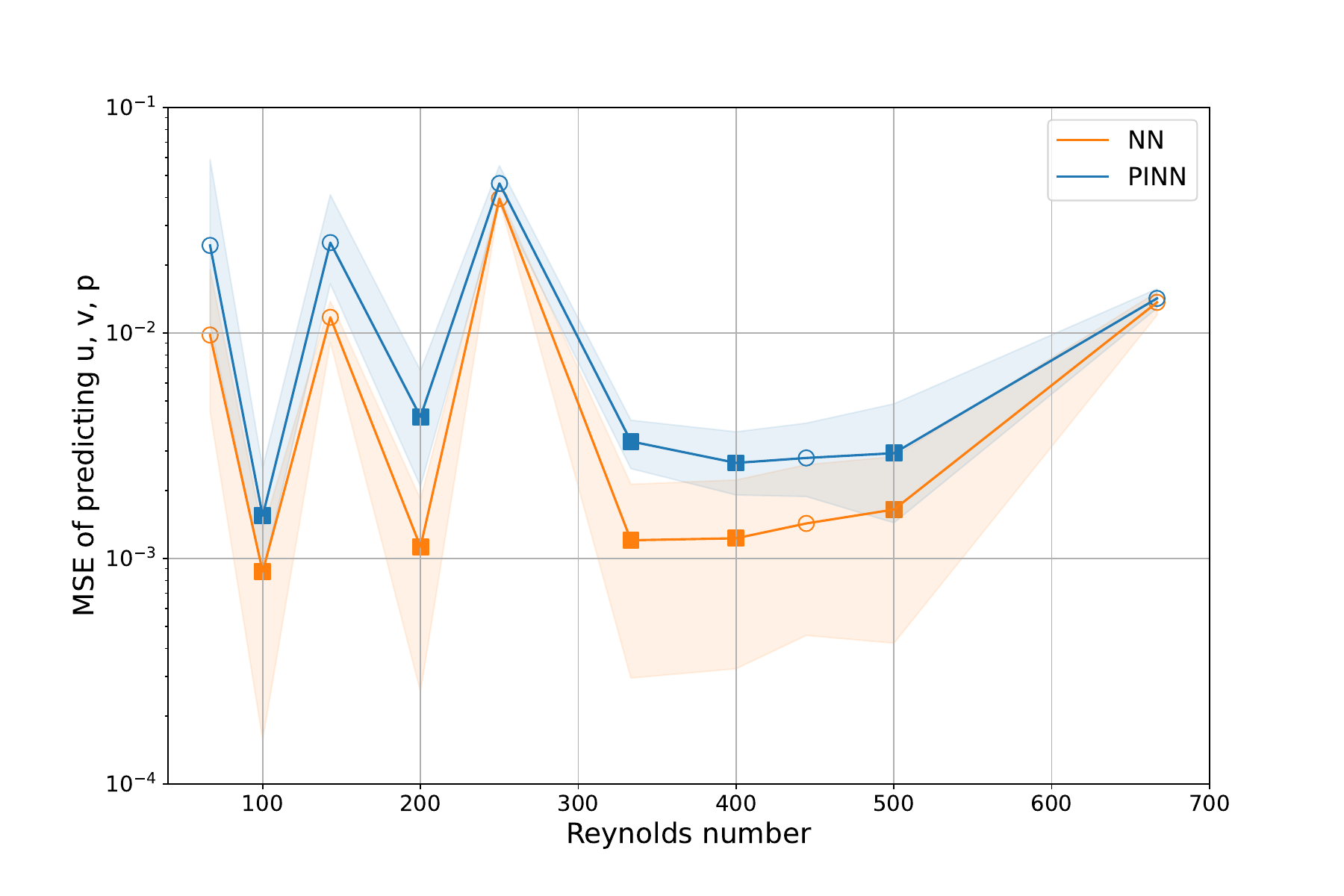}
    \caption{A comparison on the performance of the PINNs and the fully connected Neural Networks on predicting the velocity and pressure fields of the flow past cylinder problem for a range of Re numbers. Models are trained on velocity and pressure data generated from numerical simulation of the flow at: $\mathrm{Re = [100, 200, 333.\bar{3}, 400, 500]}$ (specified with filled square markers). The test set includes data from $\mathrm{Re = [66.\bar{6}, 142.8, 250, 444.\bar{4}, 666.\bar{6}]}$ as well (specified with empty circle markers). }
    \label{fig:Er_vs_Re_uvp}
\end{figure}

\textbf{Velocities and pressure}. Fig. \ref{fig:Er_vs_Re_uvp} shows the Mean Square Error (MSE) of the PINN and a NN model over a range of Re numbers. As can be observed, the NN slightly outperforms the PINN in predicting the velocities and pressures, in both the training and test sets (Re numbers present in the training set are represented with marked points). Both models predict the unseen flow conditions with one order of magnitude error higher than the corresponding error for seen flow conditions. Although these two models have the same architecture, it is apparent that constraining the PINN model with the governing PDEs has led to a reduction in accuracy in predicting the labels. Another effect of PDEs constraints is the robustness of the PINN model regarding different initializations of its adjustable parameters. Both models have been trained for $\mathrm{10}$ random weight initializations and the results in Fig. \ref{fig:Er_vs_Re_uvp} show a substantially larger standard variation for the NN than for the PINN.

\begin{figure}[h]
    \centering
    \begin{subfigure}{0.49\linewidth}
      \includegraphics[width=1\textwidth]{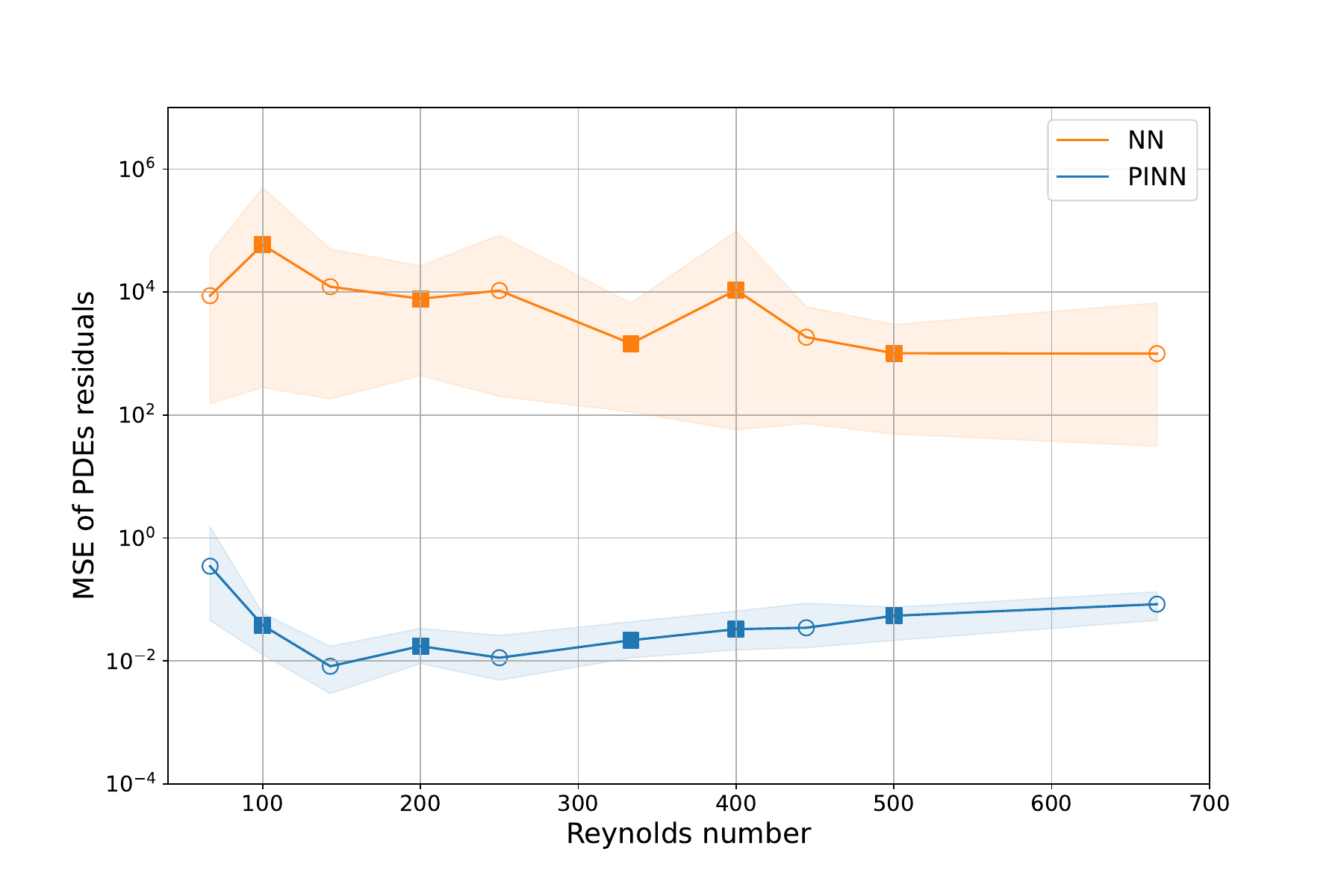}
      \caption{MSE of PDEs residuals vs Re.}\label{subfig:Er_vs_Re_PDEs}
    \end{subfigure}
    \begin{subfigure}{0.49\linewidth}
      \includegraphics[width=1\textwidth]{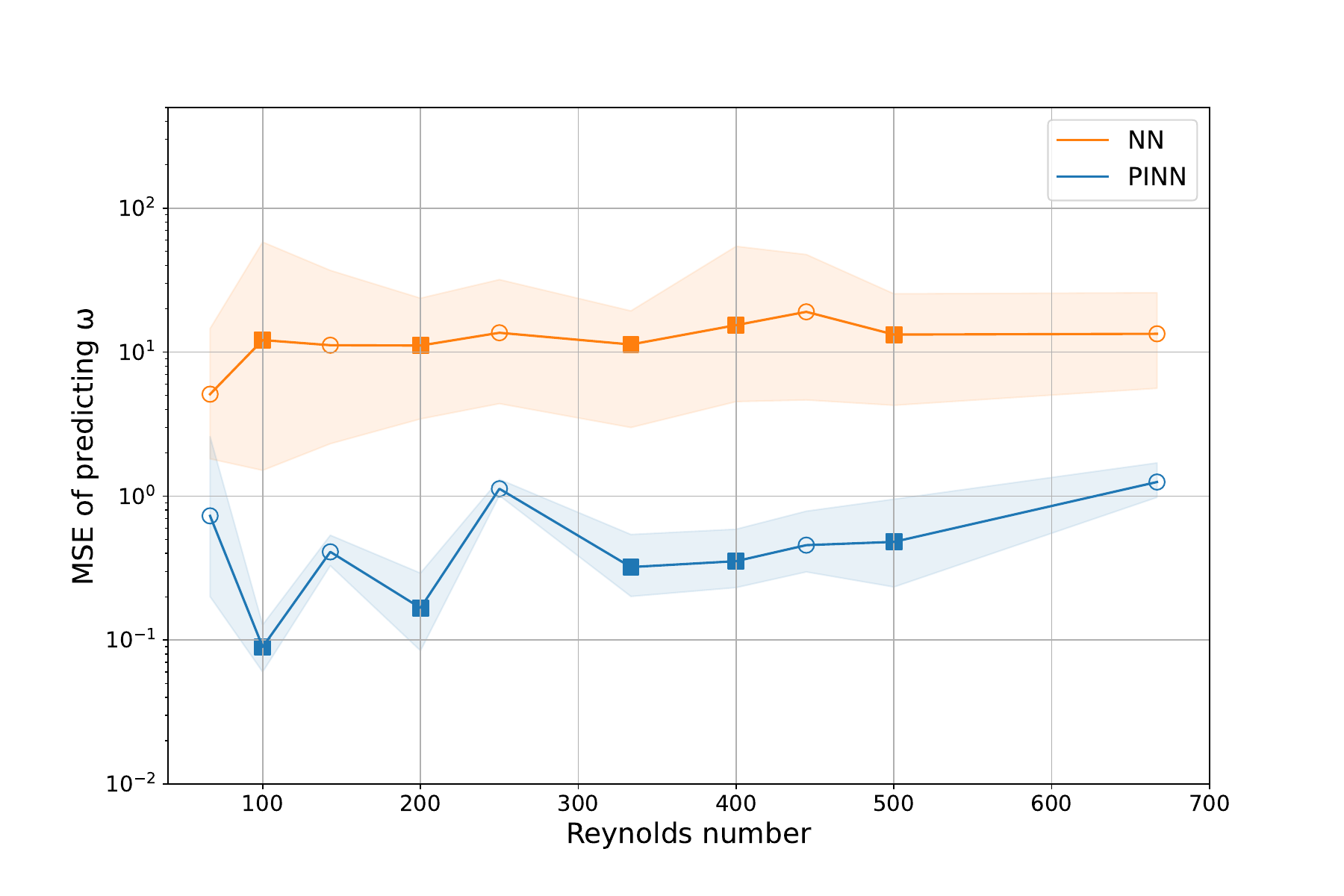} 
      \caption{MSE in prediction of vorticity vs Re.}\label{subfig:Er_vs_Re_vorticity}
    \end{subfigure}    
    % \subcaptionbox{\label{fig:Er_vs_Re_PDEs}}{
    % \includegraphics[width=0.49\textwidth]{resources/Er_vs_Re_PDEs.PDF}}
    % \subcaptionbox{\label{fig:Er_vs_Re_vorticity}}{
    % \includegraphics[width=0.49\textwidth]{resources/Er_vs_Re_vorticity.PDF}}    
    \caption{A comparison on the performance of PINNs and the NN models on (a) satisfying PDEs residuals, (b) vorticity field predictions for different Re numbers. Models are trained on velocity and pressure data generated from numerical simulation of the flow at: $\mathrm{Re = [100, 200, 333.\bar{3}, 400, 500]}$ (specified with filled square markers). Test set includes data from $\mathrm{Re} = [66.\bar{6}, 142.8, 250, 444.\bar{4}, 666.\bar{6}]$ as well (specified with empty circle markers).}
    \label{fig:Er_vs_Re_PDEW}
\end{figure}

\textbf{PDEs residual and Vorticity}. In fluid flow applications, other than predicting the velocities and pressure field, there is also interest in predicting their gradients for the purpose of computing forces on bodies and mass or heat fluxes. In Fig. \ref{subfig:Er_vs_Re_PDEs}, the errors of the PDEs residuals and vorticity are shown. While PINN's error on PDEs residuals is in the order of $10^{-1}$, NN predictions are associated with a large error (larger than $10^{3}$) in satisfying the Navier-Stokes equations. This error is more evident when predicting gradients through automatic differentiation. Fig. \ref{subfig:Er_vs_Re_vorticity} shows the error in prediction of the vorticities (note that these models are trained with velocities and pressure as model outputs and none of them have seen training data with vorticity labels). 

\begin{figure}[ht]
    \centering
    \begin{tabular}{cccc}
    \centering
    & Exact & PINN's Prediction & Absolute Error\\
    \multirow{-7}{*}{\rotatebox[origin=c]{90}{$Re = 100$}}&\includegraphics[width=0.3\textwidth]{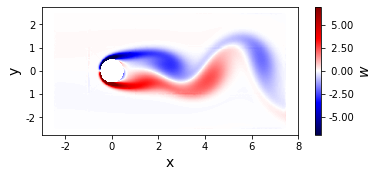}&\includegraphics[width=0.3\textwidth]{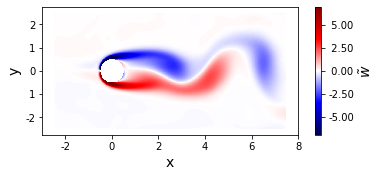}&\includegraphics[width=0.3\textwidth]{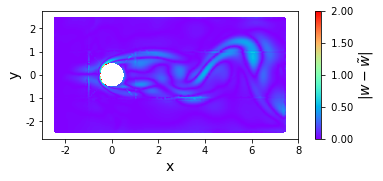}\\
    
     \multirow{-7}{*}{\rotatebox[origin=c]{90}{$Re = 500$}}&\includegraphics[width=0.3\textwidth]{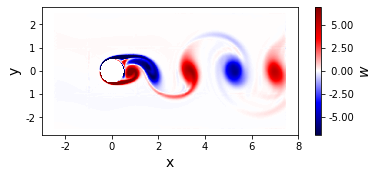}&\includegraphics[width=0.3\textwidth]{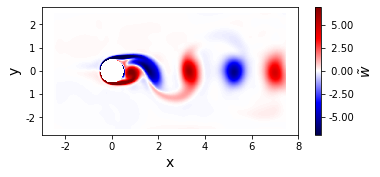}&\includegraphics[width=0.3\textwidth]{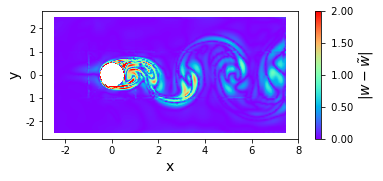}\\

    \multirow{-7}{*}{\rotatebox[origin=c]{90}{$Re = 250$}}&\includegraphics[width=0.3\textwidth]{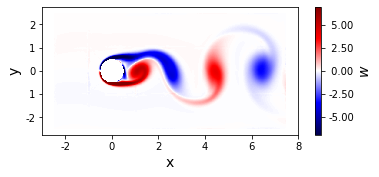}&\includegraphics[width=0.3\textwidth]{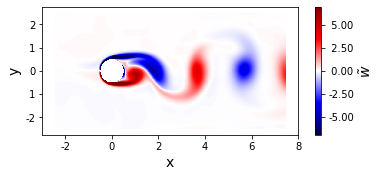}&\includegraphics[width=0.3\textwidth]{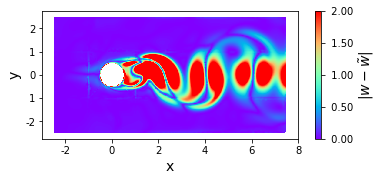}\\
    \end{tabular}
    \caption{A snapshot of PINN's Predictions of vorticity (curl of the velocity field) for fluid flow around horizontal cylinder. The model is trained on velocity and pressure data at $\mathrm{Re = [100, 200, 333.\bar{3}, 400, 500]}$, the vorticity is computed with automatic differentiation ($\omega = \nabla \times V$, $V=u\vec{i}+v\vec{j}$). }
    \label{fig:PINN_vorticities}
\end{figure}

\begin{figure}[ht]
    \centering
    \begin{tabular}{cccc}
    \centering
    & Exact & NN's Prediction & Absolute Error\\
    \multirow{-7}{*}{\rotatebox[origin=c]{90}{$Re = 100$}}&\includegraphics[width=0.3\textwidth]{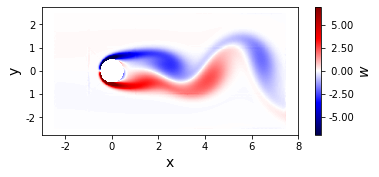}&\includegraphics[width=0.3\textwidth]{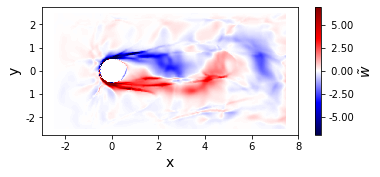}&\includegraphics[width=0.3\textwidth]{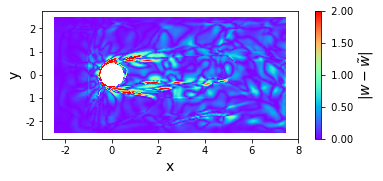}\\
    
     \multirow{-7}{*}{\rotatebox[origin=c]{90}{$Re = 500$}}&\includegraphics[width=0.3\textwidth]{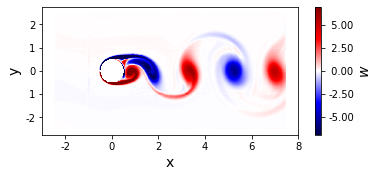}&\includegraphics[width=0.3\textwidth]{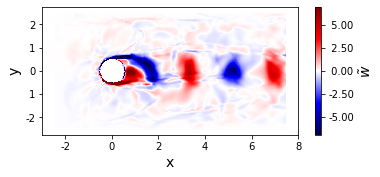}&\includegraphics[width=0.3\textwidth]{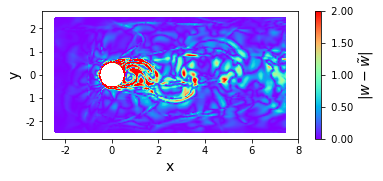}\\

    \multirow{-7}{*}{\rotatebox[origin=c]{90}{$Re = 250$}}&\includegraphics[width=0.3\textwidth]{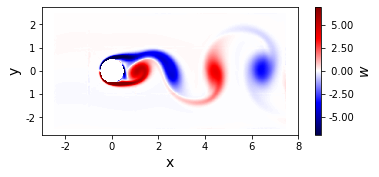}&\includegraphics[width=0.3\textwidth]{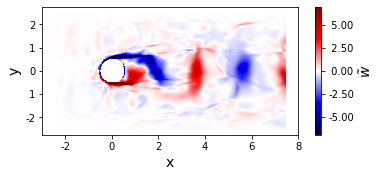}&\includegraphics[width=0.3\textwidth]{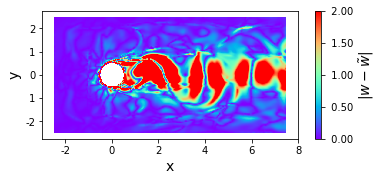}\\
     
    \end{tabular}
    \caption{A snapshot of Neural Network's Predictions of vorticity (curl of the velocity field) for fluid flow around horizontal cylinder. The model is trained on velocity and pressure data at $\mathrm{Re = [100, 200, 333.\bar{3}, 400, 500]}$, the vorticity is computed with automatic differentiation ($\omega = \nabla \times V$, $V=u\vec{i}+v\vec{j}$). }
    
    \label{fig:NN_vorticities}
\end{figure}

The PDE constraints make the PINN favorable in predicting the vorticity fields. PINN's vorticity predictions Fig. \ref{fig:PINN_vorticities} show realistic flow patterns. For seen flow conditions ($\mathrm{Re} = [100, 500]$), the largest absolute errors are apparent around the area of the cylinder and around the rotating bodies of fluid. This error increases with increasing the Re number for seen Re numbers. However, for an unseen flow condition ($\mathrm{Re} = 250$), this error is drastically larger. The major source seems to be a time shift in predictions. The same behavior is visible in the predictions by the NN model (Fig. \ref{fig:NN_vorticities}). However, for this case, predictions are associated with small discrepancies in the flow pattern.

\begin{figure}[ht]
    \centering
    \begin{subfigure}{0.49\linewidth}
      \includegraphics[width=1\textwidth]{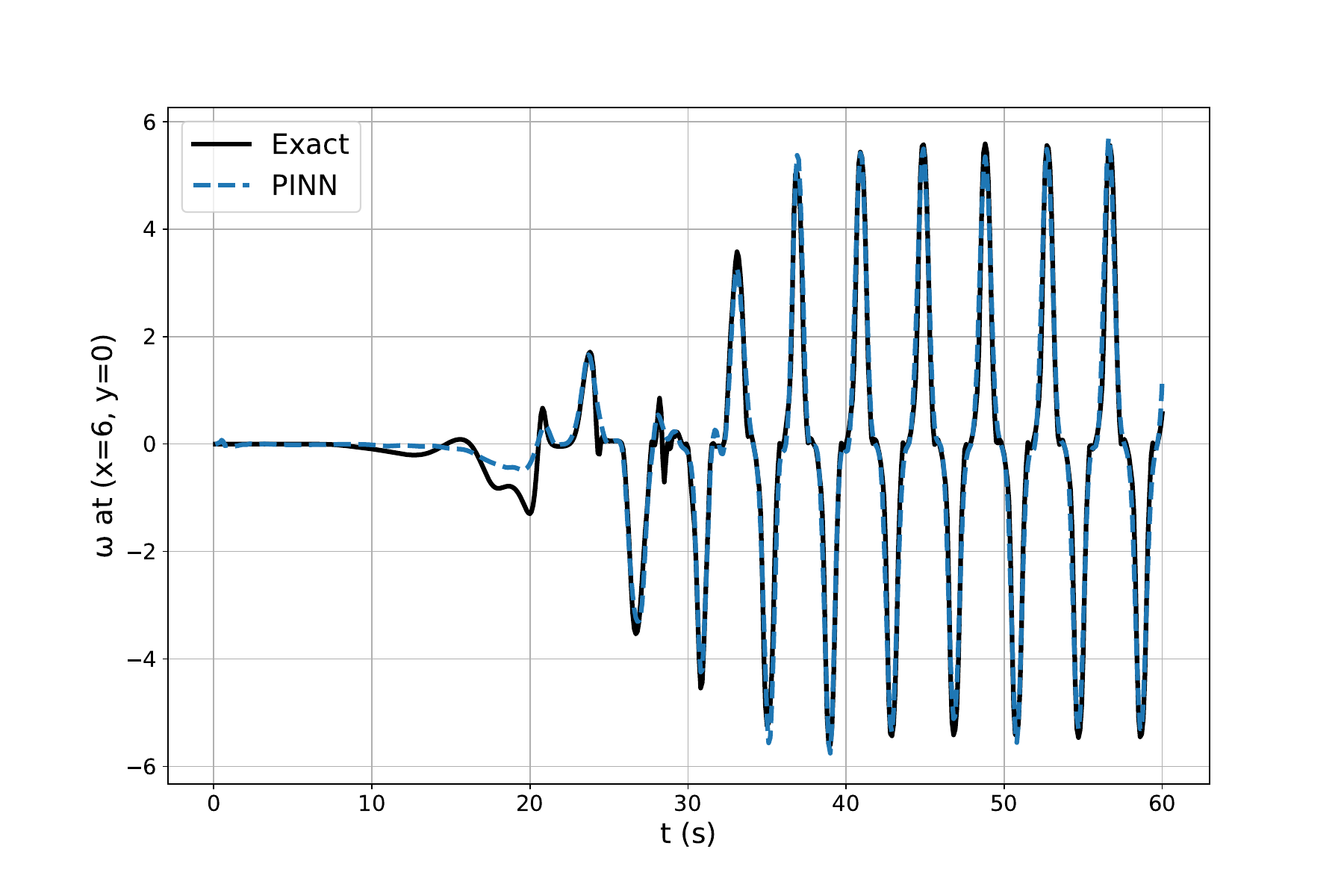}
      \caption{$\mathrm{Re}=500$ (train)}\label{subfig:Re500_2e_3_vorticity_x6_y0}
    \end{subfigure}
    \begin{subfigure}{0.49\linewidth}
      \includegraphics[width=1\textwidth]{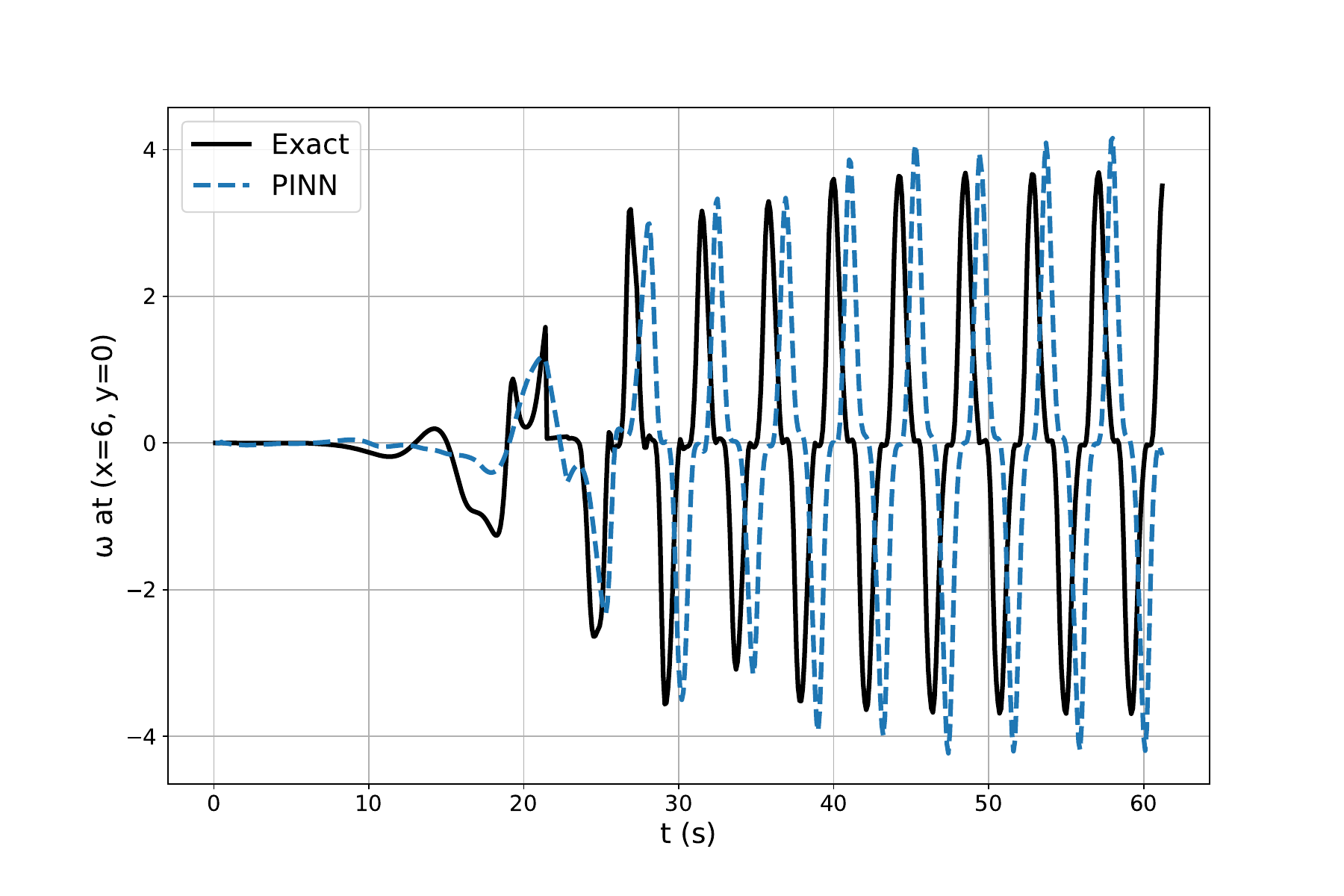} 
      \caption{$\mathrm{Re}=250$ (test)}\label{subfig:Re250_4e_3_vorticity_x6_y0}
    \end{subfigure}    
    \caption{PINN's prediction versus exact time series of vorticity at point (x=6, y=0). While predictions for seen Re numbers (a) are in line with exact data, for unseen Re numbers (b), the prediction show a time shift with exact vorticity data.}
    \label{fig:exact_vs_PINN_vorticity}
\end{figure}

The prevailing time shift in predictions at the unseen Reynolds numbers is apparent in absolute error plots of the vorticity field in Fig. \ref{fig:PINN_vorticities} and \ref{fig:NN_vorticities}. Time series plots of vorticity at local points (e.g., $x=6, y=0$) depicts these time shifts in details (Fig. \ref{fig:exact_vs_PINN_vorticity}). While the predicted vorticity time series for the seen Reynolds numbers is in line with the exact time series (e.g., $\mathrm{Re}=500$, Fig. \ref{subfig:Re500_2e_3_vorticity_x6_y0}), it is associated with a time shift for unseen Reynolds numbers (e.g., $\mathrm{Re}=250$, Fig. \ref{subfig:Re250_4e_3_vorticity_x6_y0}). Also note that the amount of time shift stays constant over time.

% \clearpage

\begin{figure}[ht]
    \centering
    \begin{subfigure}{0.49\linewidth}
      \includegraphics[width=1\textwidth]{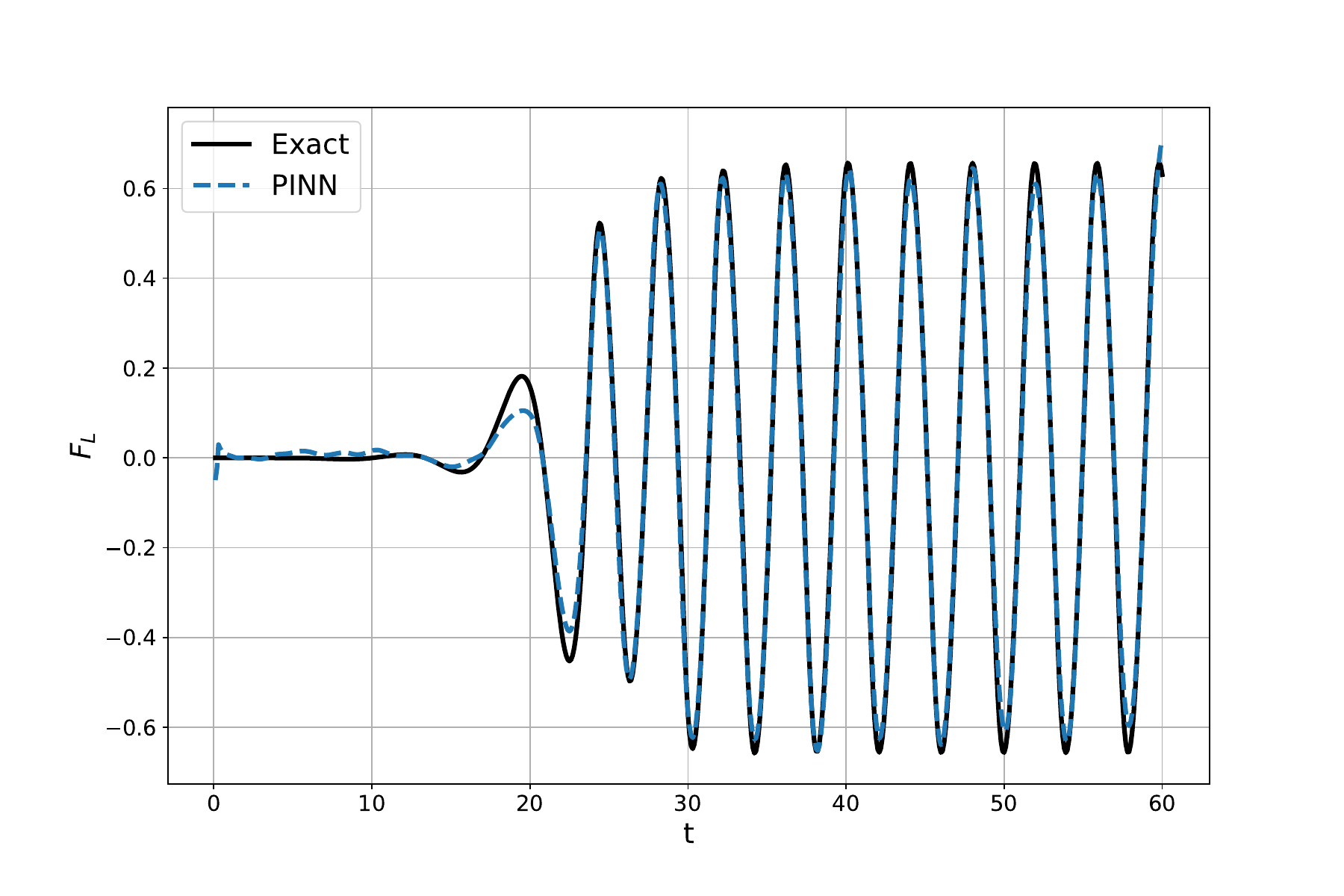}
      \caption{$\mathrm{Re}=500$ (train)}\label{subfig:Re500_2e_3_lift}
    \end{subfigure}
    \begin{subfigure}{0.49\linewidth}
      \includegraphics[width=1\textwidth]{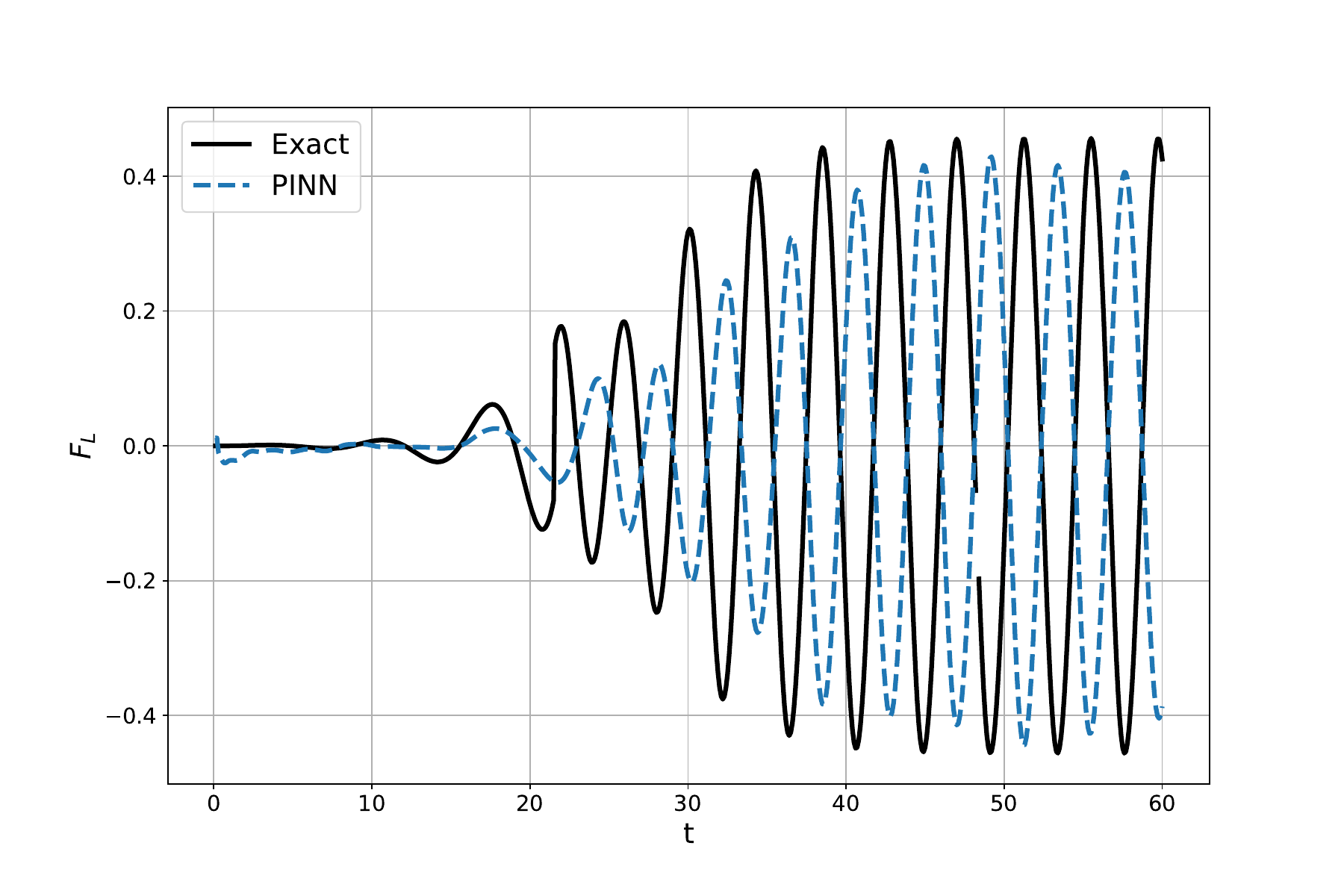} 
      \caption{$\mathrm{Re}=250$ (test)}\label{subfig:Re250_4e_3_lift}
    \end{subfigure}    
    \caption{PINN's prediction versus exact time series of Lift force ($F_{L}$) on the cylinder (appendix \ref{appendix:b}). While predictions for seen Re number (a) are in line with exact data, for unseen Re numbers (b) the prediction show a time shift with exact Lift force computed on data from numerical simulations.}
    \label{fig:exact_vs_PINN_lift}
\end{figure}
\textbf{Lift Force}. Similar to the vorticity field predictions, the predicted frequency and amplitude of lift force on the cylinder (perpendicular to the direction of the inlet flow) for seen Re numbers (Fig. \ref{subfig:Re500_2e_3_lift}) is in good agreement with the exact ones. However, for unseen Re numbers (Fig. \ref{subfig:Re250_4e_3_lift}) the predictions are associated with a time shift in oscillations. Details of lift force computations are provided in appendix \ref{appendix:b}.

\textbf{Sampling of training points}. Both PINN and NN models are flexible regarding the strategies for sampling labeled or residual data points. Here, we have sampled both type of points from uniform distributions in both the spatio-temporal and parameter domains. This implies that regions in the domain that have less importance regarding the embedded information have the same contribution in the training dataset as the important regions (e.g., the wake in the wake after the cylinder). In addition, a uniform sampling in the parameter domain similarly results in giving the same amount of attention for flows with low and high Re numbers. A visible result of this indifference in sampling is shown in Fig. \ref{subfig:Er_vs_Re_vorticity} as higher Re numbers are associated with higher error in vorticity predictions of PINN. This could have been avoided by a smarter sampling strategy favouring regions of high importance.

% \textbf{A paragraph about Re numbers}\\

% \clearpage
\section{Conclusions}\label{Conclusions}
% \textbf{Contributions.} 
In this paper, we propose the application of Physics-Informed Neural Networks for the task of learning the general solution function of parametric Navier-Stokes Equations. This solution is general as it also provides estimates of flow fields for Reynolds number that are not available during training of the PINN. In this approach, the parameter of interest interest (the Reynolds number) is considered as a direct input to the model. We tried to bypass the limitations of PINNs, when solving this set of highly non-linear PDEs by training the model on a limited set of available solutions generated by a classical numerical method for several instantiations of the parameter of interest. We investigated this setup for the classical problem of flow past a 2D cylinder. In this setup, the trained PINN model not only allows for fast prediction of the solution for unseen parameters but it also makes sure that conservational laws of mass and momentum are preserved.

% To learn the general solution function of 2D Navier-Stokes for the range of Reynolds numbers (from 100 to 500), we trained a PINN model and a unconstrained conventional Neural Network (NN) model. Training data consists of velocities and pressure labels for spatio-temporal points, the PINN model is also trained on randomly sampled residual points (no-labels). 
Our results shows that although constraining the PINN model with governing PDEs results in slight reduction of accuracy, it is beneficial in predicting realistic solutions that are satisfying conservational laws. While the unconstrained NN model is more accurate on predicting velocities and pressure, it results in drastic errors in prediction of gradients using automatic differentiation. Our results show that although the PINN model is not trained on any vorticity data, it is predicting comparably accurate vorticity fields on seen Reynolds numbers. For the unseen Re numbers, the major source of absolute error between predictions and the exact solution approximated with classical numerical methods is a time shift in predictions. Although the source of the time shift in predictions remains unknown and requires further investigations, it is less important as the ultimate periodic solution is of major interest in most fluid applications. Broadly speaking, we view our contributions as first steps towards fast approximation of flow fields for unseen flow conditions given that a limited set of solutions are available. Although this approach is still associated with considerable errors in predictions for unseen conditions, it results in valuable models for specific applications in which speed of predictions (with tolerable error margin) is the priority. 

% \textbf{Limitations and future work.} 
Our problem setup for learning parametric solutions of navier-stokes equations is based on providing simulated data for PINNs during training, provision of such data helps the PINN model in its optimization. However, it opens up several question on the number of required simulations, as well as how to sample across the parameter domain. In addition, the simulated data might be leveraged for better sampling of the residual points, however this work does not investigate such insights for better sampling and does not provide answer for the previously mentioned  questions. The current problem setup covers the first $60$ seconds of the flow until the flow reaches to its periodic oscillating state, however this might seem unnecessary for cases in which only predicting the final periodic behaviour is important. Therefore, further investigations are required for finding an optimal way to deal with time in initial value problems.

\section*{Acknowledgments}
% This research project is supported by AI4B.io consortia.
We would like to acknowledge helpful discussions and feedbacks provided by Prof. Henk Noorman
, Cees Haringa, and Jiangtao Lu. This work is supported by the AI4b.io program, a collaboration between TU Delft and dsm-firmenich, and is fully funded by dsm-firmenich and the RVO (Rijksdienst voor Ondernemend Nederland).

% \textbf{issues}
% \begin{itemize}
%     \item Introduce feature map and its necessity  in the  introduction section
%     \item here I have assumed readers know PINNs are fast in inference
% \end{itemize}
\clearpage
% \section{Appendix}
\appendix
\section{CFD Solver and computational grid.}\label{appendix:a}
\begin{figure}[h]
    \centering
    \begin{subfigure}{0.45\linewidth}
      \includegraphics[width=1\textwidth]{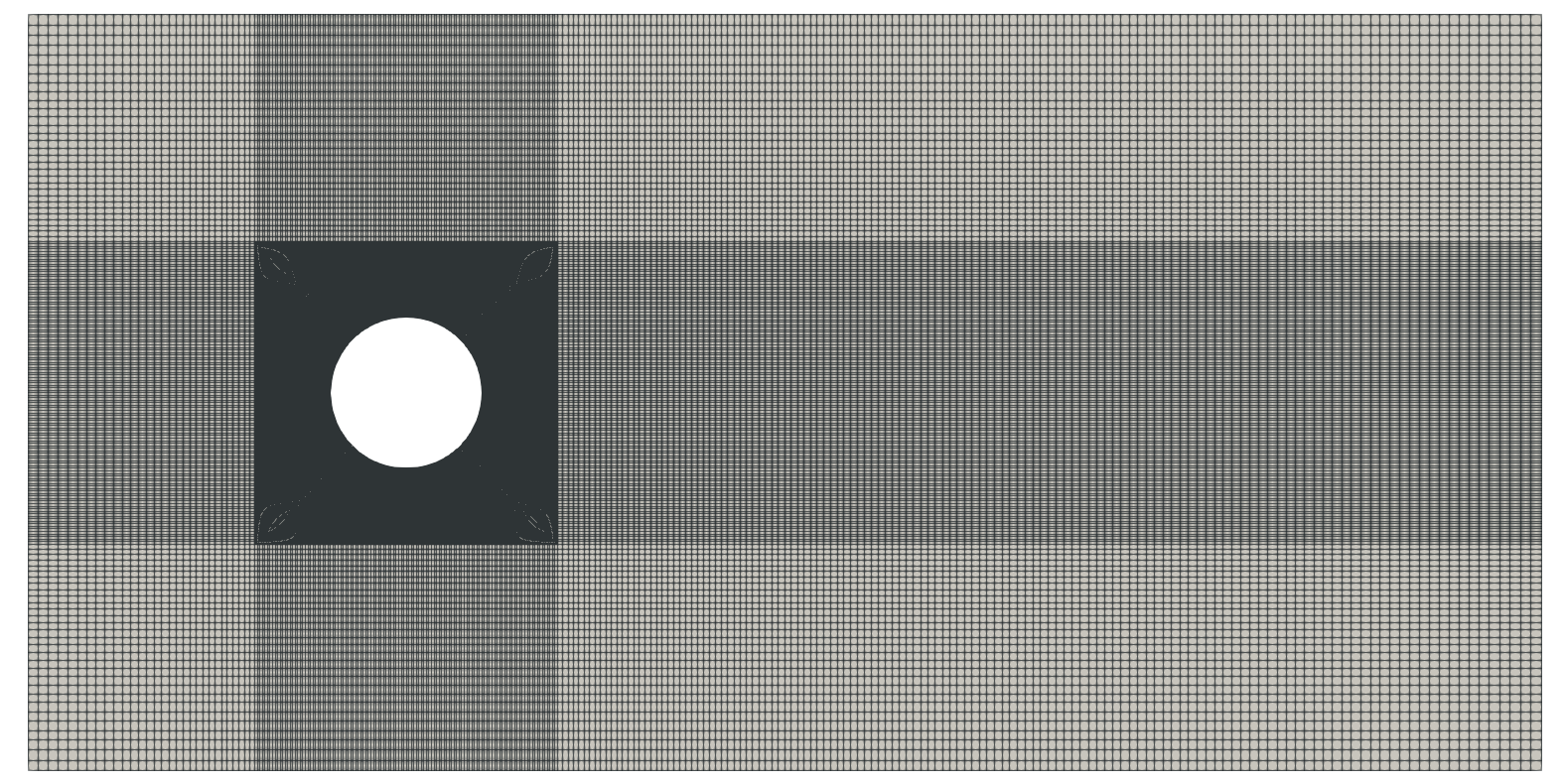}
      \caption{}%\label{subfig:mesh0}
    \end{subfigure}
    \hspace{0.05\linewidth}
    \begin{subfigure}{0.45\linewidth}
      \includegraphics[width=0.5\textwidth]{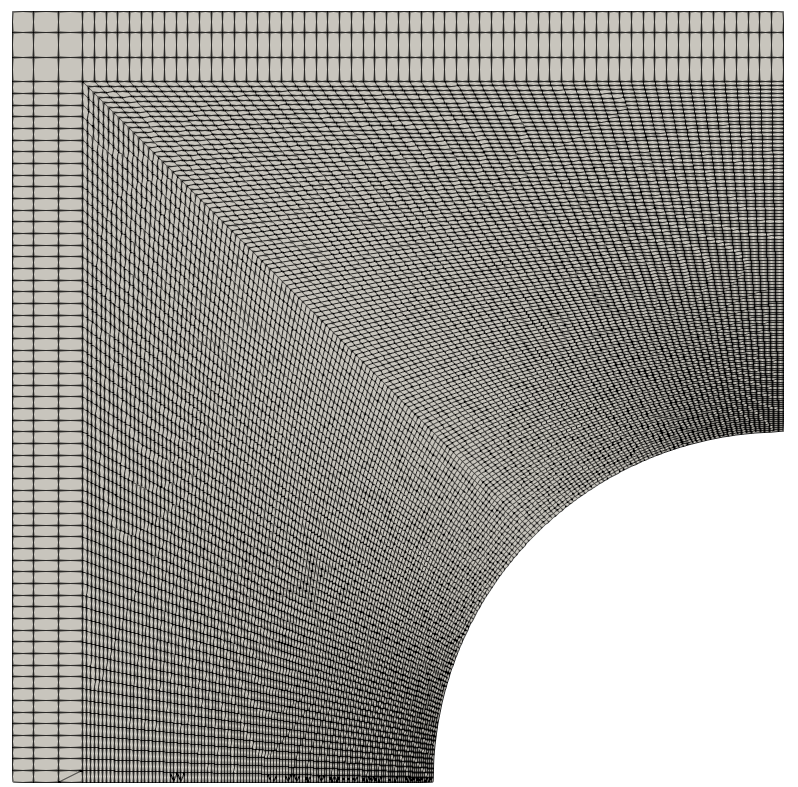}
      \caption{}%\label{subfig:mesh1}
    \end{subfigure}    
    \caption{a) Structured grid used in OpenFoam simulations, (b) a closer view on the upper left side of the cylinder}
    \label{fig:mesh}
\end{figure}

To generate the required training data for different values of Re number in the classical in-compressible flow past cylinder problem we use OpenFoam's icoFoam solver. The structured computational grid consists of $94\mathrm{k}$ hexahedral cells (Fig. \ref{fig:mesh}).

\section{Lift force computations.}\label{appendix:b}
In the flow past cylinder problem, the cylinder experience a set of forces due to motion in fluid that results in pressure and velocity gradients. Here we focus on lift, a force that acts perpendicular to the inlet flow direction. We use Eq. \ref{eq:lift_force} to compute this force.

\begin{equation}
    F_ {L} = \oint [-pn_ {y} + 2Re^ {-1} \frac{\partial v}{\partial y}  n_ {y}  +  Re^ {-1}  ( \frac{\partial u}{\partial y}  +  \frac{\partial v}{\partial x}  )  n_ {x}  ]ds
    \label{eq:lift_force}
\end{equation}

In this equation $ds$ is the length of an element on the cylinder's edge and $n_{x}$ and $n_{y}$ are components of surface normal on $ds$. Similar to \cite{raissi2018hidden}, we use trapezoidal rule for integration to approximate this integral.

% \bibliographystyle{elsarticle-num} 
\bibliographystyle{elsarticle-num-names} 

\bibliography{manual}





% \end{thebibliography}
\end{document}